\documentclass[journal]{IEEEtran}

\usepackage[utf8]{inputenc}
\usepackage[T1]{fontenc}
\usepackage{lmodern}
\usepackage{graphicx}
\usepackage{amsmath,amssymb,amsfonts}
\usepackage{siunitx} 
\usepackage{cite}    
\usepackage{url}
\usepackage{xcolor}
\usepackage[pagewise]{lineno}
\usepackage{hyperref}
\usepackage{doi}

\graphicspath{{figures/}}

\usepackage{filecontents}

\title{Demonstration of High-Performance Ultra-Wide Bandgap SrSnO$_3$ Top-Gated MOSFETs}

\author{Junghyun~Koo, Weideng~Sun, Donghwan~Kim, Hongseung~Lee, Chengyu~Zhu, Kiyoung~Lee, Hagyoul~Bae, Bharat~Jalan, and~Gang~Qiu
        
\thanks{This work was supported by the Nano \& Material Technology Development Program through the National Research Foundation of Korea (NRF) funded by the Ministry of Science and ICT under award RS-2024-00460372. MBE growth (D.K. and B.J.) was supported by the National Science Foundation (NSF) under award DMR-2306273. Film growth used instrumentation funded by AFOSR DURIP awards FA9550-18-1-0294 and FA9550-23-1-0085. (Corresponding author: Gang Qiu.)}
\thanks{J. Koo, W. Sun, C. Zhu, and G. Qiu are with the Department of Electrical and Computer Engineering, University of Minnesota, Minneapolis, MN 55455, USA (e-mail: gqiu@umn.edu). D. Kim and B. Jalan are with the Department of Chemical Engineering and Materials Science, University of Minnesota, Minneapolis, MN 55455, USA. H. Lee and H. Bae are with the Department of Electronic Engineering, Jeonbuk National University, Jeonju 54896, Republic of Korea. K. Lee is with the Department of Materials Science and Engineering, Hongik University, Seoul 04066, Republic of Korea.}}

\begin{document}
\maketitle

\begin{abstract}
We report the demonstration of high-performance top-gated metal–oxide–semiconductor field-effect transistors (MOSFETs) based on the ultra-wide bandgap perovskite oxide SrSnO$_3$ (SSO). Using hybrid molecular beam epitaxy-grown SSO channels and ALD-deposited HfO$_2$ gate dielectrics, the devices exhibit field-effect mobility exceeding $65\,\mathrm{cm^2/V\cdot s}$, an on-state current up to $194\,\mathrm{mA/mm}$, an on/off current ratio above \num{1e8}, and a contact resistance of $0.66\,\Omega\!\cdot\!\mathrm{mm}$. The devices also show a near-ideal subthreshold slope of $68\,\mathrm{mV/dec}$ and negligible hysteresis, indicating a high-quality dielectric/semiconductor interface. These results establish SrSnO$_3$ as a promising ultra-wide bandgap oxide semiconductor platform for high-performance power electronic applications.
\end{abstract}


\section{Introduction}
Ultra-wide bandgap (UWBG) semiconductors are critical for next-generation power electronics as they can endure high-voltage operations. Beyond the incumbent SiC and GaN, emerging UWBG material platforms such as $\beta$-Ga$_2$O$_3$\cite{Moser-IEDL-38-775, Green-IEDL-37-902, Chabak-IEDL-39-67}, diamond\cite{Kobayashi-Carbon-235-120024, Matsumoto-SR-6-31585, Zhang-Carbon-175-615} and AlN\cite{Chettri-JPDAP-58-035104} have shown promise with outstanding electrical properties. Recently, there has been an increasing interest in complex-oxide perovskites (ABO$_3$) as promising candidates for next-generation oxide electronics, due to their isostructural lattices and oxide–oxide compatibility that allows flexible heterostructure design\cite{Prakash-AMI-15-1900479, Nunn-JMR-36-4846}. Their robust thermal and chemical stability also offers unique advantages for potential harsh environment operations. Among perovskite oxide materials, alkaline-earth stannates have a dispersive Sn–5s conduction band and weak electron-phonon interaction, supporting light electron mass and high room-temperature electron mobilty\cite{Liu-SA-10-eadq7892, Truttmann-CP-4-241}, and are therefore promising for electronic applications.

Here, we demonstrate high-mobility top-gated enhancement-mode MOSFETs using La-doped SrSnO$_3$ (SSO). SSO has a large direct bandgap of \SI{4.1}{eV}\cite{Chaganti-IEDL-39-1381, Liu-SA-10-eadq7892}. Using hybrid molecular beam epitaxy (hMBE)-grown SSO thin films, we fabricated top-gated MOSFETs that exhibit on/off ratios $\mathrm{>}\,10^{8}$, field-effect mobility ($\mu_{\mathrm{FE}}$) exceeding $65\,\mathrm{cm^2/V\cdot s}$, an on-state current of $194\,\mathrm{mA/mm}$, subthreshold swing (SS) of $68\,\mathrm{mV/dec}$, and negligible hysteresis. Employing metal-oxide-semiconductor gate stacks enables better gate electrostatic control, higher drive current, and better interface stability, surpassing the performance of previously reported Schottky-gated SSO metal-semiconductor field-effect transistors (MESFETs)\cite{Chaganti-IEDL-41-1428, Wen-IEDL-42-74}.

\section{Material Growth and Transport Characterization}

\begin{figure}[t]\vspace{-2mm}
    \centering
    \includegraphics[width=0.9\columnwidth]{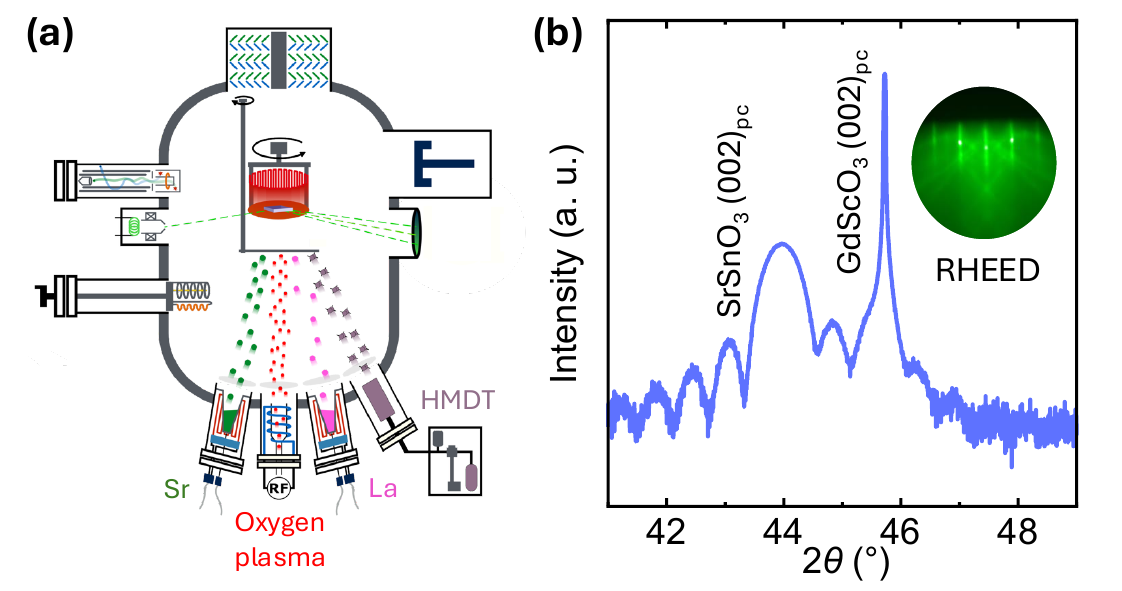}
    \vspace{-2mm}
    \caption{(a) The schematic of the radical-based hybrid MBE (hMBE) configuration for {SrSnO}$_3$ growth. (b) The High-resolution XRD 2$\theta$–$\omega$ scan of the sample. The inset shows the RHEED pattern of a pseudo-cubic phase La:{SrSnO}$_3$ film grown on pseudo-cubic phase {GdScO}$_3$ (110), acquired with the {GdScO}$_3$ $[1\bar{1}0]_{\mathrm{orth}}$ direction.}
    \label{fig:MatCharac}
    \vspace{-2mm}
\end{figure}

SSO epitaxial films were synthesized by a radical-based hMBE approach. A schematic of the hMBE is shown in Fig.\,\ref{fig:MatCharac}(a). SSO films with a typical thickness of \SIrange{12}{15}{nm} were epitaxially grown on insulating GdScO$_3$\,(GSO) (110) substrates at \SI{700}{\celsius}, and $n$-type doping was introduced by La. Additional procedural details for hMBE growth of stannate perovskites are available elsewhere\cite{Truttmann-APL-115-152103, Chaganti-IEDL-39-1381, Wang-ACSAMI-10-43802, Liu-SA-10-eadq7892}. The high‐resolution X-ray diffraction (HRXRD) 2$\theta$–$\omega$ scan in Fig.\,\ref{fig:MatCharac}(b) confirms a single‐phase, strain-stabilized tetragonal SSO\cite{Wang-ACSAMI-10-43802, Truttmann-ACSAEM-3-1127}, and the clear Laue oscillations further indicate film uniformity with high structural quality\cite{Miller-ZNB-77-313, Truttmann-ACSAEM-3-1127, Wang-ACSAMI-10-43802, Liu-SA-10-eadq7892}. The inset in Fig.\,\ref{fig:MatCharac}(b) shows a representative post-growth reflection high-energy electron diffraction (RHEED) pattern.

\begin{figure}[t]\vspace{-2mm}
    \centering
    \includegraphics[width=0.8\columnwidth]{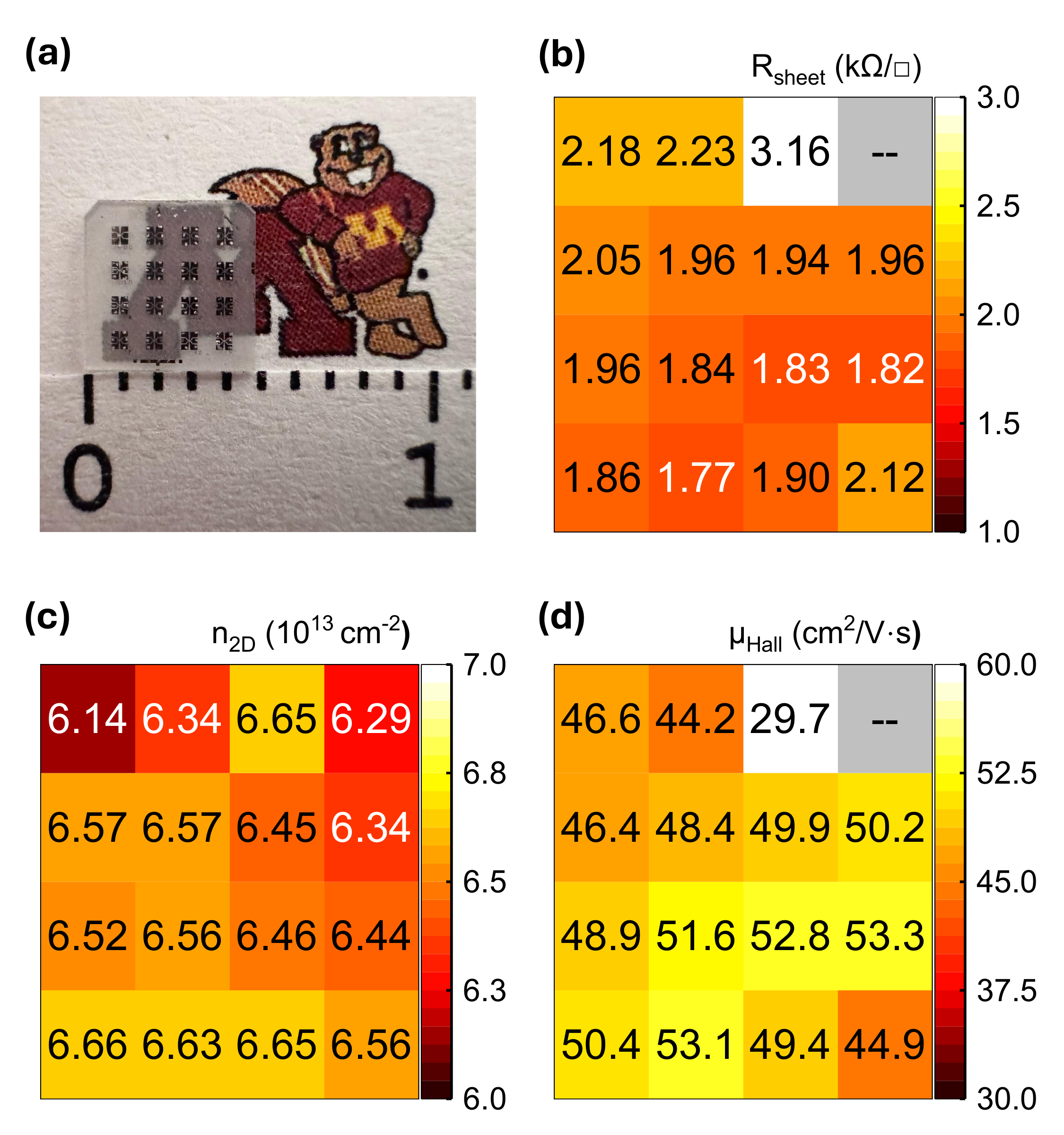}
    \vspace{-2mm}
    \caption{(a) An optical image of a $5\times5$\,mm 15-nm-thick SSO/GSO (110) wafer with a millimeter ruler, illustrating the wafer size and optical transparency. (b-d) Room-temperature sheet resistance $R_{\mathrm{sheet}}$, sheet carrier density $n_{2\mathrm{D}}$, and  Hall mobility $\mu_{\mathrm{Hall}}$ heatmaps from the $4\times4$ Hall bar array. One device in the upper left corner could not be measured due to damage during fabrication.}
    \label{fig:Hall}
    \vspace{-2mm}
\end{figure}

To assess the spatial uniformity and electrical transport properties of the film, an array of $4\times4$ six-terminal Hall bar devices was patterned on a $5\times5$\,mm 15-nm-thick SSO/GSO (110) substrate [Fig.\,\ref{fig:Hall}(a)]. Room-temperature Hall measurements were performed across all Hall bars (except one defective device due to fabrication failure), and the spatial distributions of room-temperature sheet resistance ($R_{\mathrm{sheet}}$), sheet carrier density ($n_{2\mathrm{D}}$), Hall mobility ($\mu_{\mathrm{Hall}}$) are summarized in Fig.\,\ref{fig:Hall}(b–d). Excluding the damaged device, the Hall mobility exhibits good uniformity across the wafer, with an average value of $48.0 \pm 5.8\,\mathrm{cm^2/V\cdot s}$. The $n_{2\mathrm{D}}$ values are on the order of $(6.49 ~\pm~ 0.15) \times10^{13}\,\mathrm{/cm^2}$, consistent with the doping concentration, suggesting shallow-level La dopants are fully ionized. The relatively small variation in both $\mu_{\mathrm{Hall}}$ and $n_{2\mathrm{D}}$ indicates good wafer-scale uniformity of the epitaxial SSO film. Further improvement in uniformity may be achieved through optimization of growth parameters and substrate rotation during hMBE growth.

\section{Device Fabrication}

Top-gated SSO MOSFETs were fabricated as described below. A 15-nm-thick SSO thin film (a separate sample from Fig.\,\ref{fig:Hall}) was first grown using the procedure stated previously, with a sheet carrier density of $9.45\times10^{12}\,\mathrm{/cm^2}$ and Hall mobility of $43.1\,\mathrm{cm^2/V\cdot s}$ as determined by van der Pauw measurements. After optical lithography, Cr/Au (50/50\,nm) was deposited for ohmic contacts. Channel mesas were then defined by SF$_6$/Ar reactive-ion etching (RIE). A gentle, low-power O$_2$ plasma clean preceded the ALD deposition of a 15\,nm HfO$_2$ gate insulator at \SI{200}{\celsius}. Finally, the Cr/Au gate was patterned and deposited. A cross-sectional schematic of the completed device structure is shown in Fig.\,\ref{fig:IVcurves}(a). The devices fabricated have a variety of channel lengths from 3 to 20\,\(\mu\)m and gate lengths of 4 to 30\,\(\mu\)m.

\section{Results and Discussion}

\begin{figure}[t]\vspace{-2mm}
    \centering
    \includegraphics[width=0.9\columnwidth]{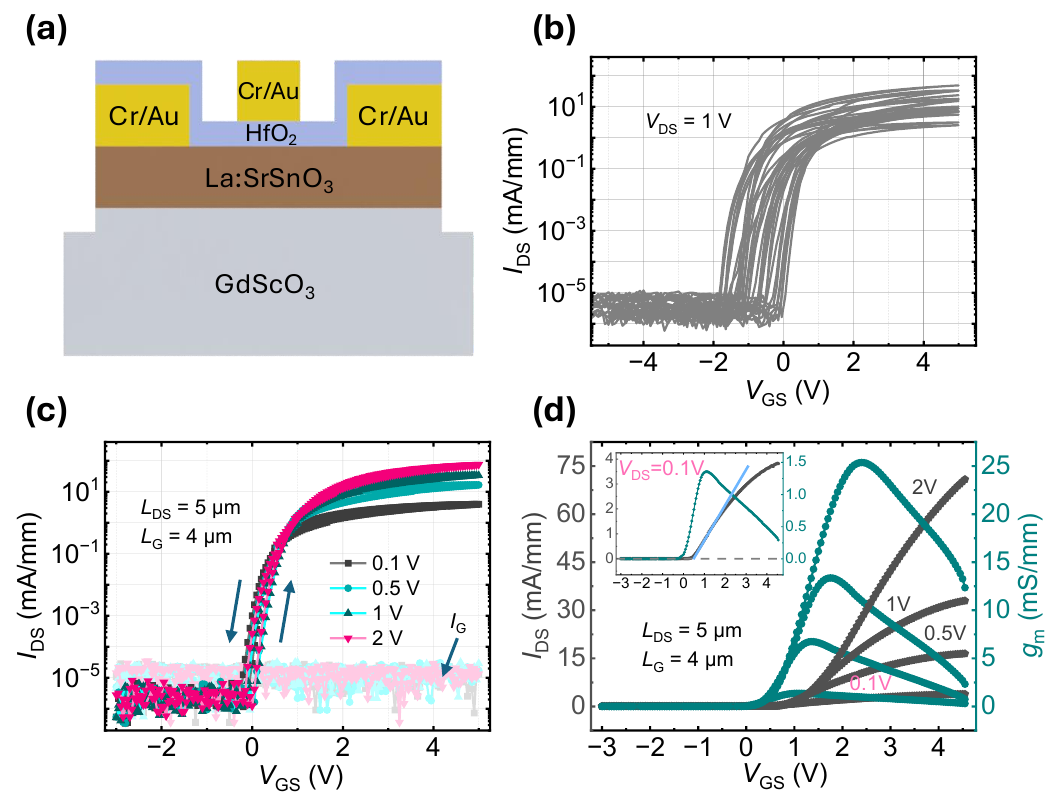}
    \vspace{-2mm}
    \caption{(a) Cross-sectional schematic of a SSO MOSFET. (b) Transfer curves of eighteen devices with different $W_{\mathrm{ch}}$, $L_\mathrm{DS}$, and $L_\mathrm{G}$. (c\,–\,d) Data from a representative device with $W_{\mathrm{ch}} = 10\,\mu$m, $L_\mathrm{DS} = 5\,\mu$m, and $L_\mathrm{G} = 4\,\mu$m: (c) Semi-log $I_\mathrm{DS}$\,–\,$V_\mathrm{GS}$ characteristics with $I_\mathrm{G}$ plotted in lighter color, (d) Linear $I_\mathrm{DS}$\,–\,$V_\mathrm{GS}$ (black, left axis) and transconductance $g_\mathrm{m}$ (green, right axis) curves. The inset shows the enlarged curves with $V_\mathrm{DS}\,=\,0.1$\,V.}
    \label{fig:IVcurves}
    \vspace{-2mm}
\end{figure}

\begin{table*}[!t]\vspace{-2mm}
    \centering
    \caption{Summary of key electrical parameters for $\mathrm{SrSnO_3}$ MOSFETs with various device geometries.}
    \label{tab:device_summary}
    \vspace{-2mm}
    \begin{tabular}{ccccccc}
    \hline
    \textbf{Device \#} & \textbf{On/Off Ratio} & \textbf{SS (mV/dec)} & \textbf{\boldmath$\mu_\mathrm{FE}$ (cm$^{2}$/V$\cdot$s)} & \textbf{\boldmath$I_\mathrm{on}$ (mA/mm)} & \textbf{\boldmath$L_{\mathrm{DS}}$ ($\mu$m)} & \textbf{\boldmath$L_{\mathrm{G}}$ ($\mu$m)} \\
    \hline
    1 & $4.0\times10^{7}$ & 97.33  & 31.4 & 150  & 5  & 4  \\
    2 & $1.7\times10^{8}$ & 68.82  & 65.9 & 194  & 5  & 4  \\
    3 & $2.3\times10^{7}$ & 164.18 & 46.2 & 150  & 5  & 6  \\
    4 & $1.0\times10^{8}$ & 68.49  & 62.7 & 90   & 5  & 6  \\
    5 & $1.5\times10^{7}$ & 113.76 & 120  & 145  & 5  & 6  \\
    6 & $2.0\times10^{7}$ & 95.31  & 54.2 & 34.3 & 10 & 10 \\
    7 & $5.0\times10^{6}$ & 98.76  & 31.8 & 47.2 & 10 & 15 \\
    \hline
    \end{tabular}
    \vspace{-2mm}
\end{table*}

The electrical performance of the as-fabricated SSO MOSFETs was measured in the dark at room temperature. Fig.\,\ref{fig:IVcurves}(b) shows the transfer curves (dual sweep) of 18 SSO MOSFETs measured at $V_\mathrm{DS}\,=\,1$\,V, with an average threshold voltage ($V_{\mathrm{th}}$) of $-0.07 \pm 0.68$~V. Several devices exhibit enhancement-mode behavior with positive threshold voltage. Note that the 18 devices shown in Fig.\,\ref{fig:IVcurves}(b) have different dimensions ($L_{\mathrm{DS}}$, $W_{\mathrm{ch}}$, $L_{\mathrm{G}}$) and geometries (whether or not the gate overlaps the source/drain). This variation, along with the modest film inhomogeneity [Fig.\,\ref{fig:Hall}(b)], accounts for the minor differences in device performance.

The dual-sweep $I_{DS}$\,–\,$V_{GS}$ trace of a representative device ($L_{DS}=5\,\mu\mathrm{m}$, $L_{G}=4\,\mu\mathrm{m}$) is shown in Fig.\,\ref{fig:IVcurves}(c). The transfer curves show negligible hysteresis, indicating a low interfacial trap density. A subthreshold slope of $68\,\mathrm{mV/dec}$ over the drain-current window $I_{\mathrm{DS}}=10^{-4}$\,–\,$10^{-2}\,\mathrm{mA/mm}$ corresponds to a low interface trap density of $D_{\mathrm{it}}\approx 1.03\times10^{12}\,\mathrm{/cm^2\cdot eV}$ for the HfO$_2$/SSO interface.

From the linear $I_{\mathrm{DS}}$\,–\,$V_{\mathrm{GS}}$ (left axis) plot in Fig.\,\ref{fig:IVcurves}(d), the same representative device exhibits a positive threshold voltage of 0.5\,V, indicating enhancement-mode operation. The corresponding transconductance $g_{\mathrm{m}}=\mathrm{d}I_{\mathrm{DS}}/\mathrm{d}V_{\mathrm{GS}}$ peaks at $25.4$\,$\mathrm{mS/mm}$ for $V_{\mathrm{DS}}=$\,$2$\,V. A $\mu_{\mathrm{FE}}$ of $65.9\,\mathrm{cm^2/V\cdot s}$ is extracted from the transfer curve in the linear region ($V_{DS}=0.1$\,V)\cite{Sze-SDPT-2012}. The increase in $g_{\mathrm{m}}$ with $V_{\mathrm{GS}}$ near the threshold can be explained by the increasing accumulation carriers that screen the impurity scattering; at a higher overdrive voltages, the stronger gate-induced effective vertical field $E_{\mathrm{eff}}$ attracts carriers closer to the HfO$_2$/SSO interface, promoting surface scattering that decreases mobility.

\begin{figure}[t]\vspace{-2mm}
    \centering
    \includegraphics[width=0.9\columnwidth]{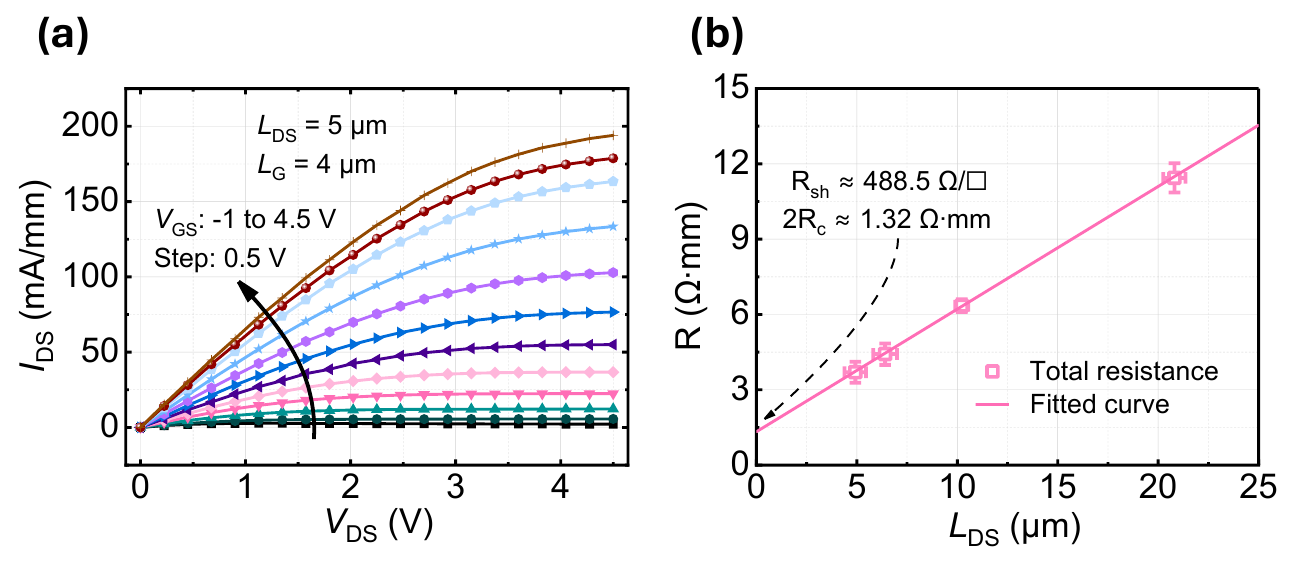}
    \vspace{-2mm}
    \caption{(a) $I_\mathrm{DS}$\,–\,$V_\mathrm{DS}$ curves for $V_\mathrm{GS}$ from $-1$ to $4.5$\,V in $0.5$\,V increments. (b) The TLM plot for Cr/Au source/drain contacts: width-normalized total resistance $R_{\mathrm{tot}}W$ versus contact spacing $L$ with vertical (measurement) and horizontal (S/D-spacing) error bars. Linear fitting gives a sheet resistance of $R_{\mathrm{sh}}\approx 488.5\,\Omega/\square$ and an intercept of $2R_{\mathrm{c}}\approx 1.32\,\Omega\!\cdot\!\mathrm{mm}$, corresponding to $R_{\mathrm{c}}\approx 0.66\,\Omega\!\cdot\!\mathrm{mm}$.}

    \label{fig:TLM}
    \vspace{-2mm}
\end{figure}

The output characteristics in Fig.\,\ref{fig:TLM}(a) show a large on-state current of $194\,\mathrm{mA/mm}$ at $V_{\mathrm{GS}}\,=\,V_{\mathrm{DS}}\,=\,4.5\,\mathrm{V}$, despite the relatively long channel length of $5\,\mu\mathrm{m}$. The on-state resistance of $14.7 \,\Omega\!\cdot\!\mathrm{mm}$ is extracted from the low-$V_{\mathrm{DS}}$ linear regime of the output curve under gate bias of 4.5 V [Fig.\,\ref{fig:TLM}(a)], which corresponds to a specific on-resistance of $0.74 \,\mathrm{m}\Omega\!\cdot\!\mathrm{cm^2}$. The high on-state current and the linear $I_{\mathrm{DS}}$–$V_{\mathrm{DS}}$ behavior in the low-$V_{\mathrm{DS}}$ regime indicate efficient carrier injection and low contact resistance at the metal/SSO interface. 

To quantitatively evaluate the contact resistance, transfer-length method (TLM) measurements were performed using Cr/Au (50/50\,nm) contacts fabricated on a separate SSO sample under identical growth and processing conditions, as shown in Fig.\,\ref{fig:TLM}(b). The width-normalized total resistance was measured for devices with contact spacings ranging from $L=5$ to $21\,\mu\mathrm{m}$ and plotted as a function of $L$. Linear fitting yields a sheet resistance of $R_{\mathrm{sh}}\approx 488.5\,\Omega/\square$ and an intercept of $2R_{\mathrm{c}}\approx 1.32\,\Omega\!\cdot\!\mathrm{mm}$, corresponding to a contact resistance of $R_{\mathrm{c}}\approx 0.66\,\Omega\!\cdot\!\mathrm{mm}$. This low contact resistance is comparable to or lower than that in other perovskite oxide MOSFETs, including CaSnO$_3$ devices reported in Ref.\,\cite{Sun-AElM-e00459}.

\begin{figure}[t]\vspace{-2mm}
    \centering
    \includegraphics[width=0.9\columnwidth]{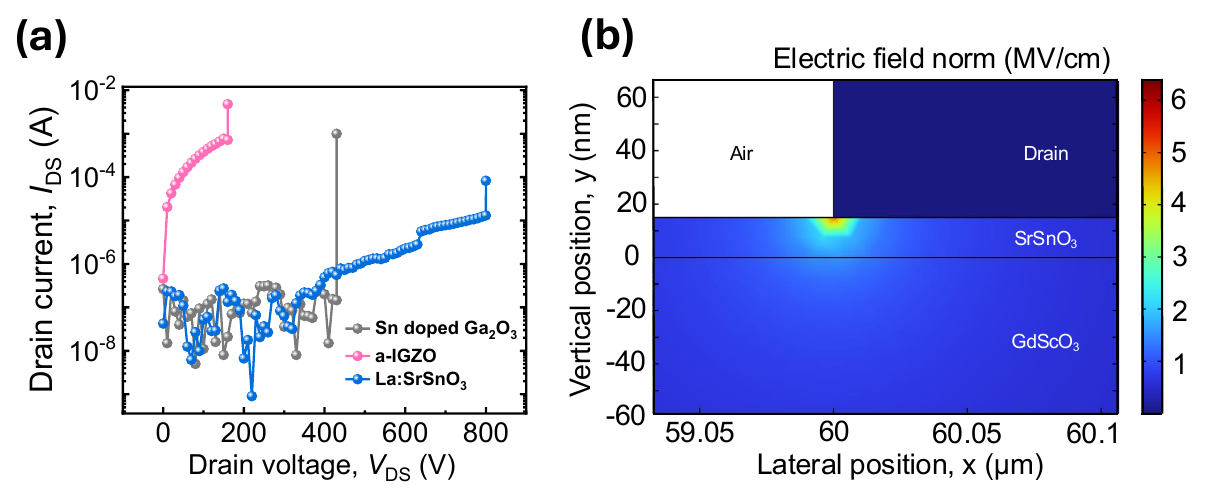}
    \vspace{-2mm}
    \caption{(a) Breakdown voltage comparison between SSO, IGZO, and Sn-doped Ga$_2$O$_3$ measured from the same two-terminal device geometry. A breakdown voltage of 800\,V is observed in the SSO device. (b) COMSOL-simulated electric field distribution under 800\,V bias, corresponding to a breakdown electric field of 6.4\,MV/cm.}
    \label{fig:Breakdown}
    \vspace{-2mm}
\end{figure}

We note that the excellent electrical performance is consistently observed across multiple devices, as summarized in Table\,\ref{tab:device_summary}, demonstrating the reproducibility and robustness of the SSO MOSFETs. On average, these devices show $\mu_{\mathrm{FE}}$ on the order of $60\,\mathrm{cm^2/V\cdot s}$, large on/off ratios of $\sim 10^{8}$,  and low SS of $100\,\mathrm{mV/dec}$. Notably, a peak $\mu_{\mathrm{FE}}$ of $120\,\mathrm{cm^2/V\cdot s}$ is observed in the best-performing device among the measured set.

To further evaluate the suitability of SSO for power electronic applications, breakdown voltage was measured using a two-terminal device structure, with a source/drain spacing of $60\,\mu\mathrm{m}$. A breakdown voltage of $800\,\mathrm{V}$ was observed, as shown in Fig.\,\ref{fig:Breakdown}(a). For comparison, other common oxide semiconductors were tested using the same device structure and testing protocol, and SSO shows a breakdown voltage $2\times$ and $4\times$ higher than Sn-doped Ga$_2$O$_3$ and IGZO, respectively. To provide a numerical estimation of the SSO breakdown field, finite-element simulations were performed based on the geometry of the experimentally measured device. Under an applied bias of $800\,\mathrm{V}$, the simulated electric field distribution is shown in Fig.\,\ref{fig:Breakdown}(b), revealing a peak electric field of $6.4\,\mathrm{MV/cm}$ near the contact edge. The electric field distribution suggests that breakdown originates from field crowding near the contact edge and is therefore nearly independent of the channel length. This implies that further improvement of breakdown voltage may be achieved through optimization of device geometry for high-voltage applications.

\section{Conclusion}
We have demonstrated the first SSO MOSFET with high electrical performance. The devices exhibit high field-effect mobility, a large on/off ratio, low contact resistance, and high drive current enabled by high-quality hMBE-grown channels. Moreover, a low SS and near-hysteresis-free operations are achieved with a clean HfO$_2$/SSO interface. Our work validates SSO as an ultra-wide bandgap semiconducting perovskite oxide platform for high-performance power electronics.


\bibliographystyle{IEEEtranDOI}
\bibliography{reference.bib}



\end{document}